\documentclass[preprint2]{aastex}

\newcommand{\be}{\begin{equation}}
\newcommand{\ee}{\end{equation}}
\newcommand{\ba}{\begin{eqnarray}}
\newcommand{\ea}{\end{eqnarray}}

\slugcomment{ApJ, to be submitted}

\shorttitle{SEP fluences at 1 AU}
\shortauthors{Marsh et al.}

\begin{document}


\title{Solar Energetic Particle drifts and the energy dependence of 1 AU charge states}

%(or focus title more on corotation?)


\author{S.~Dalla\altaffilmark{1}, M.S.~Marsh\altaffilmark{2}, M.~Battarbee\altaffilmark{1}}
\affil{$^1$Jeremiah Horrocks Institute, University of Central Lancashire,
    Preston, PR1 2HE, UK}
\affil{$^2$Met Office, Exeter, EX1 3PB, UK}

% \author{C. D. Biemesderfer\altaffilmark{4,5}}
% \affil{National Optical Astronomy Observatories, Tucson, AZ 85719}
% \email{aastex-help@aas.org}

% \and

% \author{R. J. Hanisch\altaffilmark{5}}
% \affil{Space Telescope Science Institute, Baltimore, MD 21218}


% \altaffiltext{1}{Visiting Astronomer, Cerro Tololo Inter-American Observatory.
% CTIO is operated by AURA, Inc.\ under contract to the National Science
% Foundation.}
% \altaffiltext{2}{Society of Fellows, Harvard University.}
% \altaffiltext{3}{present address: Center for Astrophysics,
%     60 Garden Street, Cambridge, MA 02138}
% \altaffiltext{4}{Visiting Programmer, Space Telescope Science Institute}
% \altaffiltext{5}{Patron, Alonso's Bar and Grill}


\begin{abstract}
The event-averaged charge state of heavy ion Solar Energetic Particles (SEPs), measured at 1 AU from the Sun, typically increases with the ions' kinetic energy. The origin of this behaviour has been ascribed to processes taking place within the acceleration region.
In this paper we study the propagation through interplanetary space of SEP Fe ions, injected near the Sun with a variety of charge states that are uniformly distributed in energy, by means of a 3D test particle model. 
In our simulations, due to gradient and curvature drifts associated with the Parker spiral magnetic field, ions of different charge propagate with very different efficiencies to an observer that is not magnetically well connected to the source region. As a result we find that, for many observer locations, the 1 AU event-averaged charge state $<$$Q$$>$, as obtained from our model, displays an increase with particle energy $E$, in qualitative agreement with spacecraft observations. We conclude that drift-associated propagation is a possible explanation for the observed distribution of $<$$Q$$>$ versus $E$ in SEP events, and that the distribution measured in interplanetary space cannot be taken to represent that at injection. 
\end{abstract}



\section{Introduction}

Solar Energetic Particle (SEP) events, detected by spacecraft following flares and Coronal Mass Ejections (CMEs), are characterised by enhancements in the intensities of heavy ions, in addition to proton and electron signatures.
One of the most interesting features of in-situ observations of heavy ion SEPs is that ionic charge states show a dependence on the particles' kinetic energy \citep{Moe1999,Pop2006,Kle2007}. In some SEP enhancements, the event-averaged charge state, $<$$Q$$>$, for Fe has been reported to increase with energy from about 8 at $\sim$100 keV nucleon$^{-1}$ to about 20 at $\sim$40 MeV nucleon$^{-1}$ \citep{Pop2006}. While the absolute value of $<$$Q$$>$ for a given energy range has been found to vary considerably between events, the general trend for it to increase with energy appears to be a ubiquitous feature of SEP measurements.

The observations of energy-dependent charge states have so far been interpreted as resulting mainly from the characteristics of the acceleration process. Broadly speaking, the amount of collisional stripping taking place at the acceleration region is thought to be energy dependent, because particles of different energies spend different amounts of time within the region, for both flare and shock energisation scenarios \citep{Ost2000,Bar1999}. 
In addition, it has been pointed out that adiabatic deceleration produces a shift in the distribution of $<$$Q$$>$ versus energy \citep{Kar2005}.

\begin{figure}[t]
 \epsscale{0.8}
 \plotone{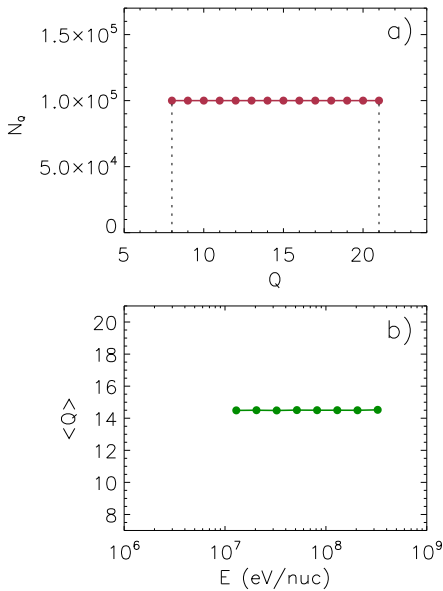}
 \caption{Parameters of the injected Fe ions: {\it(a)} number $N_Q$  of particles injected for each charge value; {\it(b)} energy distribution of average charge $<$$Q$$>$ within the population.
 \label{fig.inject_dist}}
 \end{figure}

In this paper, we point out for the first time that guiding centre drifts associated with the gradient and curvature of the interplanetary magnetic field (IMF) naturally introduce an energy dependence of observed charge states. 
Drifts have been know for many years to be important in the propagation of galactic cosmic rays (GCRs) but are typically ignored for SEPs due to their lower energies. However simulations by \cite{Mar2013} and analytical calculations by \cite{Dal2013} have shown that drift effects are not negligible for SEPs and are particularly important for heavy ions, due to their large mass-to-charge ratio.
 
Modelling of heavy ion propagation is typically carried out within spatially 1D formulations, in which the ions are assumed to remain tied to the field line on which they were originally injected, therefore neglecting drift effects (e.g.~\cite{Mas2012}). The same 1D assumption is the basis of all current interpretations of heavy ion SEP data.
Here we carry out full-orbit 3D test particle simulations of the propagation of Fe ions in the interplanetary magnetic field, and determine the average charge state that would be measured by a 1 AU observer. 
Our model contains a number of simplifying assumptions but it will nontheless allow us to give a qualitative demostration of the effect of drift on measured charge states.

\section{Simulations}\label{sec.test_simul}

We make use of a 3D full-orbit test particle code, which integrates trajectories of heavy ions through a unipolar, outward pointing, Parker spiral IMF \citep{Mar2013}. Particles are released instantaneously at a radial coordinate $r$=2 $R_{Sun}$, with $R_{Sun}$ the solar radius, over a compact region of angular extent 6$^{\circ}$$\times$6$^{\circ}$, centered at a helio\-graphic latitude $\delta$=20$^{\circ}$.
The code solves the equations of motion for each particle and drifts are included simply by the fact that a magnetic field with gradient and curvature and an electric field are specified.
The injection spectrum of the heavy ions is assumed a be a power law spectrum in energy per nucleon, with spectral index $\gamma$=1.1, in the range 10--400 MeV nucleon$^{-1}$.
Ions propagating in the IMF are assumed to experience a low level of scattering, having a mean free path $\lambda$=1 AU. 
At each scattering event the direction of the particle's velocity is reassigned randomly from a spherical distribution in velocity space. Times between scattering events used in the simulations are exponentially distributed, assuming scattering to be a Poisson process \citep{Mar2013}. 
We choose a constant mean free path (independent of rigidity) to isolate drift effects from those associated with rigidity dependent scattering.
Our previous simulations have shown that drift behaviour is only weakly dependent on the value of $\lambda$ \citep{Mar2013}. No scattering across the field is introduced in our model. 
Particles trajectories are integrated up to the final time $t_f$=100 hr.
Other parameters of the simulation are the same as in \cite{Mar2013}.

% meaning that the only transport across the field that takes place during a scattering event is of the order of the Larmor radius.

Fe ions are released into interplanetary space by the acceleration processes with a range of charge states, and the number $N_Q$ of injected ions for each $Q$ value, as well as the energy distribution of charge states, are not well known.  
Given that our aim is to study the effect of drift on the energy distribution of $<$$Q$$>$ at 1 AU in a qualitative way, we consider the following simple scenario for the particle charge at injection.  
We assume that Fe ions are released from the source region with charges between $Q$=8 and $Q$=21, with $N_Q$ constant with $Q$ (Figure \ref{fig.inject_dist}-(a)), and that there is no energy dependence of the charge states at injection, so that at all energies the average injection charge is 14.5 (Figure \ref{fig.inject_dist}-(b)).  
For each charge value, $N_Q$=10$^5$ particles are propagated in our simulation, giving a total of 1.4 million particles.

The test particle code produces as output the times of particle crossings of the 1 AU sphere, as well as the particle characteristics at that time.
This information is then used to construct intensity profiles and plots of average charge versus energy as would be detected by a 1 AU observer. All particle positions are also stored at 1 hr intervals.

\begin{figure*}
\epsscale{1.0}
\includegraphics[scale=.3]{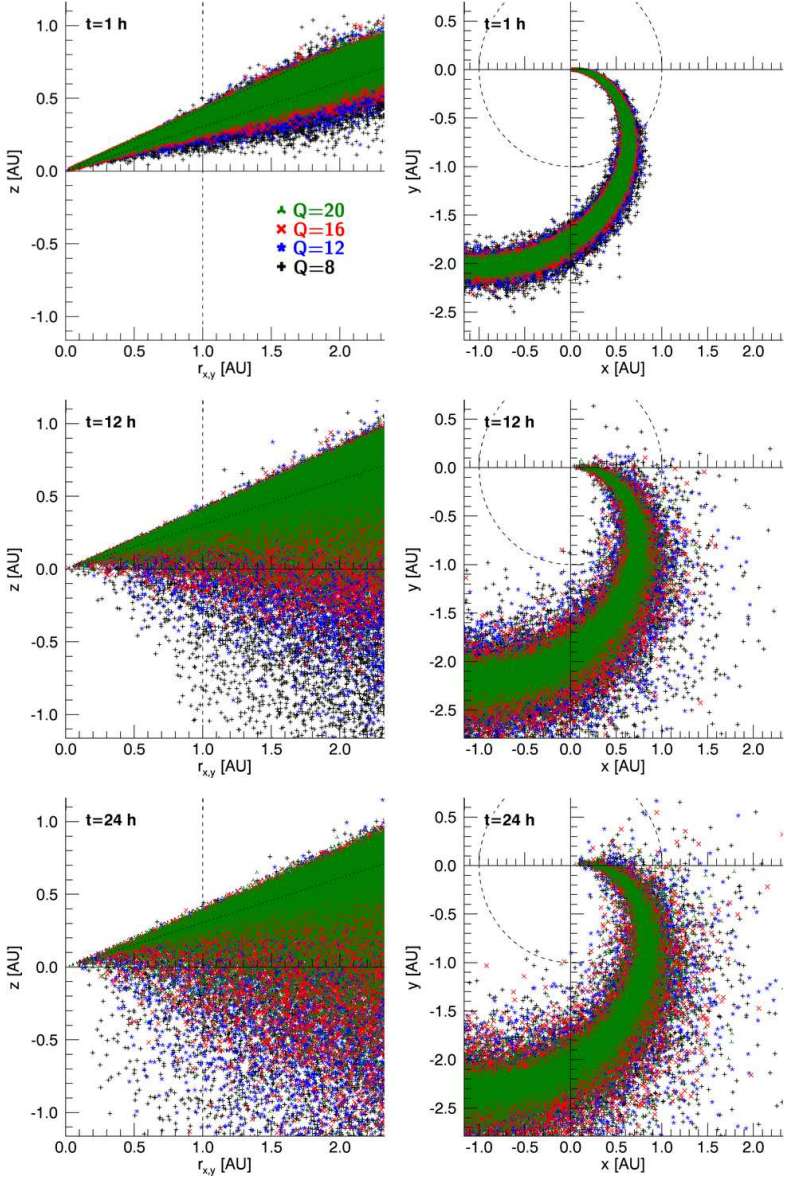} 
\caption{Locations of Fe ions of selected charges at $t$=1, 12 and 24 hr. Shown are particles for $Q$=8 ({\it black}), 12 ({\it blue}), 16 ({\it red}) and 20 ({\it green}). Left panels give $z$ vs $r_{xy}$=$(x^2+y^2)^{1/2}$  and right panels $y$ vs $x$, with only particles within 20$^{\circ}$ of the heliographic equator included in the $x$-$y$ projections. In the latter, a dashed circle at 1 AU is shown. In the left panels, the dashed line is at $r_{xy}$=1 AU and dotted lines indicate the extent of the flux tube into which injection took place. Separate contour plots of Fe ion locations for $Q$=8, 12, 16 and 20 are available as Supplementary Material.
\label{fig.scatterplot}}
\end{figure*}

\begin{figure*}
\epsscale{1.0}
\includegraphics[scale=.67]{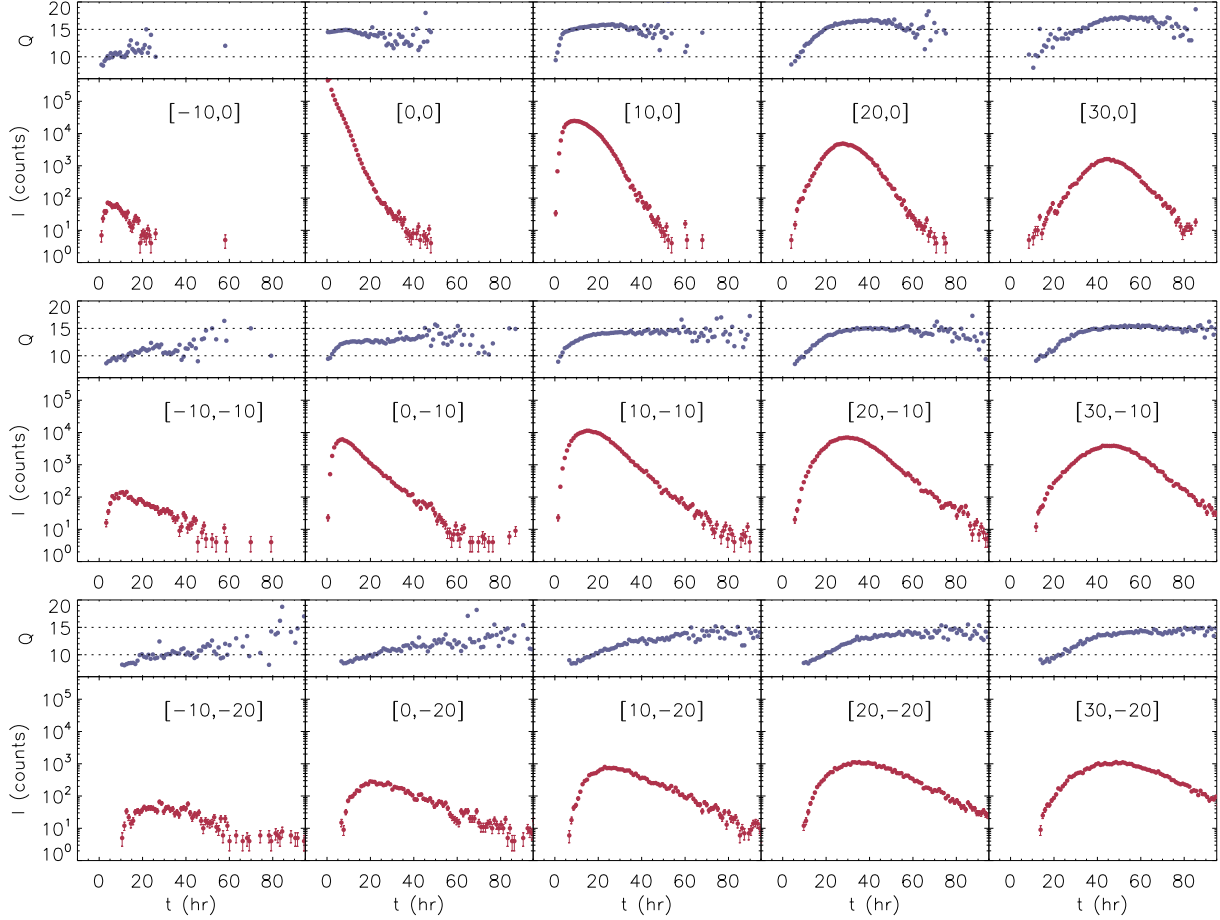} 
\caption{Fe charge $Q$ ({\it blue dots}) and counts $I$ ({\it red dots}) versus time for various observer positions at 1 AU, for energies in the range 10--80 MeV nucleon$^{-1}$. Labels in each panel give the values of $[\Delta\phi_{1AU}, \Delta\delta_{1AU}]$, where  $\Delta\phi_{1AU}$ is the heliographic longitude and $\Delta\delta_{1AU}$ the heliographic latitude of the 1 AU observer relative to the location with magnetic connection to the centre of the injection region, through a Parker spiral field line. The particle charge $Q$ is an average over the time interval over which counts are accumulated. Counts are collected over 10$^{\circ}$$\times$10$^{\circ}$ portions of the 1 AU sphere.
\label{fig.intens1}}
\end{figure*}

\begin{figure*}
\epsscale{0.95}
\includegraphics[scale=.67]{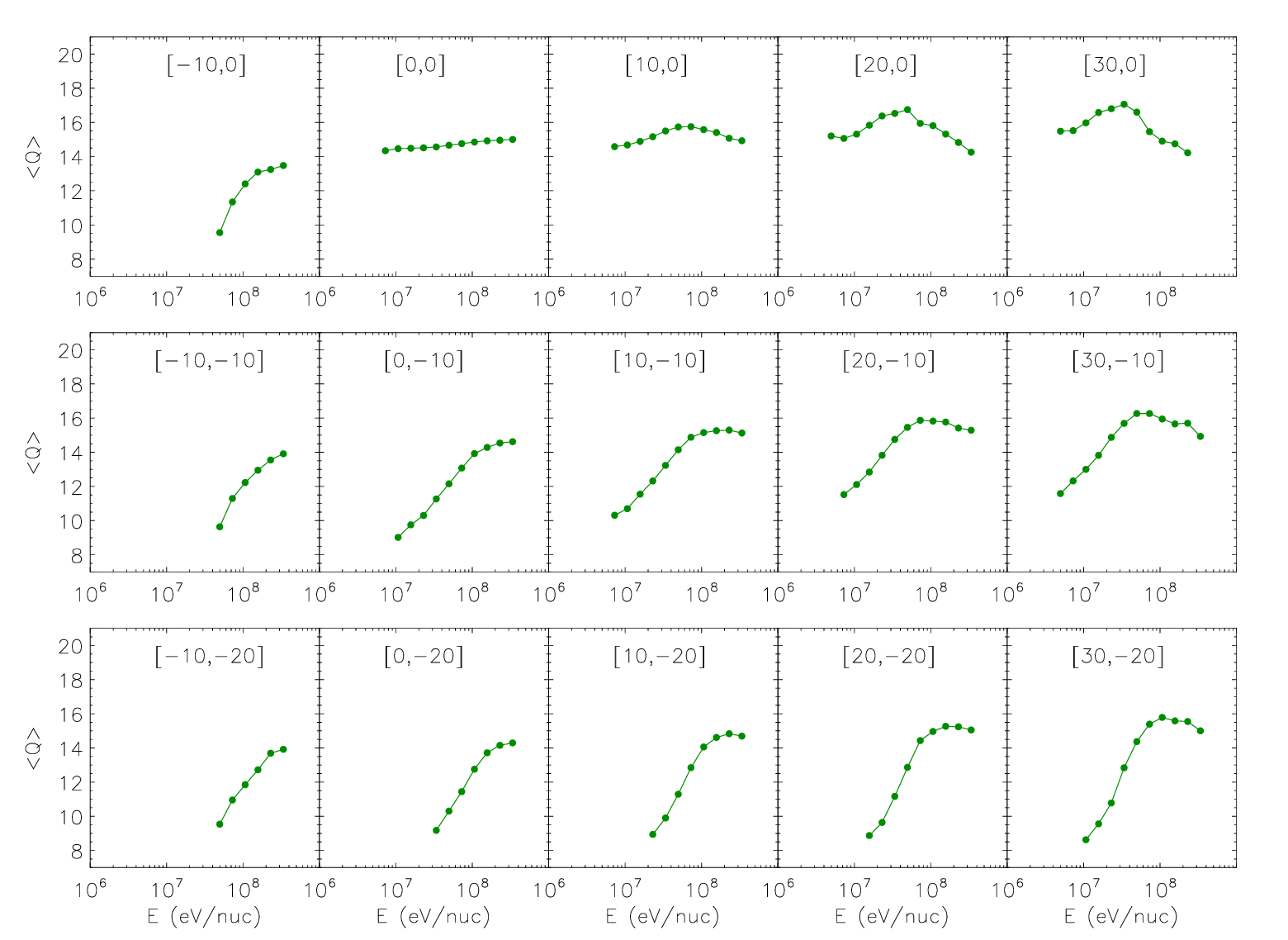}
\caption{Average Fe charge over the entire SEP event versus kinetic energy per nucleon $E$ for the same observer positions as in Figure \ref{fig.intens1}. Only energy bins with more than 200 particles are plotted.
\label{fig.q_energy}}
\end{figure*}

Figure \ref{fig.scatterplot} shows the locations of Fe ions at three times during the simulation, $t$=1, 12 and 24 hr. Only particles with charge states $Q$=8, 12, 16 and 20 are displayed in the plots, colour coded by their $Q$-value. The plots in the left column give $z$ versus $r_{xy}$=$(x^2+y^2)^{1/2}$, where $x$, $y$ and $z$ are cartesian coordinates centered at the Sun, the $z$ axis being the solar rotation axis and the $x$ and $y$ axes lying in the heliographic equatorial plane. Plots in the right column show $y$ versus $x$ for particles within 20$^{\circ}$ of the heliographic equator: this is done to emphasize the spread in longitude while minimising projection effects associated with particles that have drifted to high heliolatitudes. 

The Fe ion propagation seen in Figure \ref{fig.scatterplot} is characterised by a very efficient drift in latitude away from the injection flux tube (centered at 20$^{\circ}$ latitude) with a large number of particles having already reached the opposite hemisphere at $t$=12 hr. This propagation is clearly dependent on the charge state, with the $Q$=8 particles ({\it black symbols}) having experienced the largest drift, in agreement with analytical drift expressions \citep{Dal2013}. 
For the outward polarity used in our simulation, ions drift downwards.
In addition to the dependence on $Q$, drift velocities increase with kinetic energy as well as being dependent on pitch angle and location in the heliosphere \citep{Dal2013}. As a result, particles of a given charge are spread over a wide latitudinal range in the $z$ versus $r_{xy}$ plots. Drift velocities also have a component in the heliolongitudinal direction, making the size of the region filled by energetic particles in the $x$-$y$ projection grow over time, as the flux tubes experience corotation. 

Figure \ref{fig.intens1} shows particle parameters at 1 AU versus time, as would be detected by observers at various locations relative to the SEP source.
The larger panels of Figure \ref{fig.intens1} show counts of Fe ions versus time ({\it red dots}) in the energy range 10--80 MeV nucleon$^{-1}$ for several 1 AU observers. 
This energy range is chosen so as to represent typical SEP observations while being broad enough to ensure good statistics from our simulations.
Observer locations are specified using labels $[\Delta\phi_{1AU}, \Delta\delta_{1AU}]$, where $\Delta\phi_{1AU}$ is the 
heliographic longitude and $\Delta\delta_{1AU}$ the heliographic latitude of the observer relative to the Parker spiral field line through to the centre of the particle injection region. The panel labelled [0,0] corresponds to an observer connected to the centre of the injection region, and the other panels to less well connected observers.  Moving to the right along a row in Figure \ref{fig.intens1} one can see count profiles for observers at the same latitude and progressively more western longitudes (i.e.~source region becoming more eastern). Different rows correspond to different observer latitudes, becoming more southern as one moves downwards.
Counts are collected over 10$^{\circ}$$\times$10$^{\circ}$ portions of the 1 AU sphere.
The smaller panels of Figure \ref{fig.intens1} show the average particle charge ({\it blue dots}) within each accumulation time used to construct the counts profile.

For a well connected observer (see panel [0,0] of Figure \ref{fig.intens1}) counts peak very rapidly and decay monotonically, while the particle charge remains fairly constant over time at around the average injection charge of 14.5, for times with large number of counts in the accumulation interval.

Fe intensities are also detected by observers that are not directly connected to the injection region via magnetic field lines (all panels apart from [0,0]), indicating that the propagation is fully 3D, with significant transport across the field, due to drifts, as seen in Figure  \ref{fig.scatterplot}. The peak counts for a not well connected observer are at least an order of magnitude lower than for a well connected observer.
As the source region becomes more Eastern relative to the observer (i.e.~moving from left to right in each row of Figure \ref{fig.intens1}) the time of peak intensity shifts to a later time. Latitudinal effects are also present: the variation in peak intensity is largest for an observer at the same latitude as the source region, while at more southern latitudes peak intensities are more similar to each other. This behaviour is a result of the ions drifting downwards.
An observer for which the source region appears as Western (see panel [-10,0]) sees the event as having a smaller peak intensity and a shorter duration, and the change of the patterns of profiles with longitude
matches the typical East-West ordering \citep{Mar2015}.

Looking at the charge (blue dots in Figure \ref{fig.intens1}) of the ions contributing to each profile, one can see that at each observer's location apart from [0,0], the first arriving particles have a low charge: these are the particles with largest drift velocity due to their higher mass-to-charge ratio. Particles of higher charge tend to contribute to the later part of the profile as they drift less and arrive later.

Figure \ref{fig.q_energy} shows the event-averaged Fe charge state $<$$Q$$>$ versus kinetic energy per nucleon $E$, at the same locations as in Figure \ref{fig.intens1}.
Here one can see that when the observer is directly connected to the injection region, the 1 AU charge distribution is fairly flat, but not uniform. 
This is due to the fact that leakage out of the well connected flux tube is energy and charge dependent, with the low charges and high energies leaking most efficiently. 
When the observer is located at a different latitude from the injection region, a steep increase of charge state with energy is seen (up to 6 units over an order of magnitude in energy), with the steepness of the slope increasing with latitudinal separation.
Observers at the same latitude as the injection location and more western locations see a variation of  $<$$Q$$>$  by about two units over an order of magnitude in energy: an increase with $E$ can be seen at the lower energies and a decrease at the higher energies.

%\section{Interpretation of simulation results}

\section{Discussion}

In Section \ref{sec.test_simul} we presented SEP Fe intensity profiles at 1 AU as derived from a 3D full orbit test particle model that includes the effects of drifts. An equal number of ions was injected for each charge state from $Q$=8 to $Q$=21 and the initial energy distribution of charge states was chosen to be uniform.

Our results show that a direct magnetic connection to the source is not necessary to detect significant Fe intensities, because drift effects produce transport across the magnetic field. The propagation of these ions is therefore fully 3D and cannot be described by 1D models.

For a given observer location, our simulations show that ions of lower charge arrive first, because of their larger drift velocities, proportional to $m/q$. This prediction of our model could potentially be verified by means of sensitive instrumentation (see e.g.~\cite{Kle2007} for a discussion of the methodologies and challenges in the measurement of SEP charge states).
According to our model
the low charge (high drift) SEPs remain confined to flux tubes close to the injection one only for short times, and rapidly fill the entire heliosphere. This dilutes them, making them less likely to be detected by spacecraft instrumentation, apart from early in an event. 

We have also shown that drift-associated propagation effects significantly modify the injection charge energy distribution, for an observer not directly connected to the source location. While the energy distribution of charge states was uniform at injection, at 1 AU for most observers the event-averaged charge state $<$$Q$$>$ showed a significant increase with energy. This is in qualitative agreement with SEP Fe measurements \citep{Pop2006,Kle2007}. The slope of the measured charge distribution obtained from our simulation increases with increasing latitudinal separation between observer and source region.

The reason why drift effects modify the energy distribution of $<$$Q$$>$ is that drift velocities for non-relativistic ions are directly proportional to the product of mass-to-charge ratio by energy per nucleon \citep{Dal2013}.
To reach a not well connected observer the drift velocity needs to be sufficiently large, and, broadly speaking, at low energy per nucleon this can only be achieved by low charge particles, while at higher energy per nucleon it can be achieved by particles of higher charge. 
To this simple picture, one needs to add the effects of deceleration, including that induced by drift \citep{Dal2015}, which strongly influence the intensity profiles at a given energy and the charge energy distribution. Therefore drift effects that impact the $<$$Q$$>$ vs $E$ distribution include both transport perpendicular to the field and deceleration.

We conclude that the common assumption that the energy distribution of charge states measured at 1 AU is the same as the injection distribution, and consequently a signature of the acceleration process, is invalid.
In this paper we showed that the latter statement applies to the case of a compact injection region, so that it is immediately applicable to SEP events associated with solar flares.
For SEP events associated with CME-driven shocks, the injection region has a wide extent; we can imagine it as the sum of a number of compact injection tiles: our simulation results will apply to each individual tile and the overall intensities and charge states measured by an observer will be the superposition of the contributions of several tiles.  Within the CME-driven shock acceleration hypothesis, the efficiency of SEP injection is assumed to vary strongly along the shock front (e.g.~\citet{Rea1996, Kal1997}): therefore, applying our results to this scenario, ions experiencing 3D propagation from an injection tile e.g.~close to the nose of the shock, where acceleration is most efficient, will contribute significantly to the observables measured at locations directly connected to other parts of the shock. The measured energy charge distribution will be a contribution from many tiles and will have been \lq processed\rq\ by 3D transport, rather than being simply the injection distribution propagated parallel to the field lines, as is assumed by 1D models. 
Owing also to the latitudinal dependence of drift velocity, a non-uniform injection efficiency over the shock front will result in a complex energy distribution of charge states at 1 AU that cannot be assumed to represent the injection distribution.   

Our results have been obtained for a scattering mean free path $\lambda$=1 AU: earlier we studied 3D proton propagation for $\lambda$=0.3, 1 and 10 AU and found that drift motion across the field is only weakly dependent on the value of $\lambda$  (see Figure 3 of \citet{Mar2013}). 
In Figure 3 of \citet{Dal2016} we compared intensity profiles of Fe with $Q$=15 for $\lambda$=0.1 and 1 AU, as obtained with our 3D model. We found that the overall shape of the profiles at locations not well connected to the source region (i.e.~all locations apart from [0,0]) is not strongly affected by the choice of scattering mean free path, although the peak intensities are affected. 
We therefore conclude that a different choice for $\lambda$ in our simulation would not change the qualitative behaviour presented in Section \ref{sec.test_simul}, for not well connected locations that particles reach via drift. 

The IMF model considered in this work is that of a Parker spiral of single polarity, used in order to consider the simplest possible drift scenario. The model describes the case in which particles propagate to reach the observer within a volume of single polarity and is relevant to many actual situations in the heliosphere. The presence of two polarities and of the Heliospheric Current Sheet (HCS) will influence the drift patterns and associated particle transport. The particles that reach the HCS will experience additional current sheet drift. A full study of test particle trajectories in the presence of a HCS will be presented in a future publication (Battarbee et al, in prep., 2016).

In our simulation, we did not attempt to adjust the specific form of the $N_Q$ versus $Q$ profile and charge energy distribution at injection to obtain a 1 AU energy distribution of  $<$$Q$$>$ that matches that observed in real SEP events (both in terms of absolute values and slope). This is because there are a number of effects that are not included in our model, including a possible wide extent and long duration of the injection as well as an energy dependence at injection that results from the acceleration process  \citep{Ost2000,Bar1999}. 
In diffusive shock acceleration, turbulent trapping influences the effectiveness of ion
acceleration and particle escape in a way that depends on the mass to charge ratio \citep{Bat2011}. 
Our results show that in addition to processes taking place at the acceleration region, drift effects have a significant impact on the measured energy dependence of 1 AU charge states. 
The way in which the different processes contribute to the observed $<$$Q$$>$ versus $E$ dependence will need to be established in future work, but our simulations show that the magnitude of the drift effect can be large.

\section{Conclusions}

Our main findings can be summarised as follows:
\begin{itemize}

\item The propagation of Fe ions through the interplanetary medium needs to be modelled in 3D, due to strong drift effects. The ionic charge  influences the amount of drift experienced, so that low charges tend to arrive first at an observer not directly connected to the source region. 

\item  For a not well connected observer at latitude different from the injection latitude, drifts result in an increase of the measured 1 AU event-averaged charge state $<$$Q$$>$  with kinetic energy per nucleon $E$. 

\item The energy distribution of charge states at 1 AU can differ significantly from the injection distribution. 

\end{itemize}

In a recent paper \citep{Dal2016} we considered the 3D propagation of both Fe and O ions, and showed that differences in drift-associated transport across the magnetic field between the two species, result in a decay of the Fe/O ratio over time, similar to that observed at 1 AU in SEP events. Therefore 3D propagation drift effects described by our model can reproduce both energy dependence of charge states and Fe/O decay, two key signatures in heavy ion SEP observations that are currently ascribed to separate phenomena.

% Large Q in impulsive events?


%% Included in this acknowledgments section are examples of the
%% AASTeX hypertext markup commands. Use \url without the optional [HREF]
%% argument when you want to print the url directly in the text. Otherwise,
%% use either \url or \anchor, with the HREF as the first argument and the
%% text to be printed in the second.

\acknowledgments

This work has received funding from the UK Science and Technology Facilities Council (STFC) (grant ST/M00760X/1) and the Leverhulme Trust (grant RPG-2015-094).
SD acknowledges support from ISSI through funding for the International Team on \lq Superdiffusive transport in space plasmas and its influence
on energetic particle acceleration and propagation\rq.

\bibliographystyle{apj}
\bibliography{qenergy_biblio}

%\begin{thebibliography}{}
%\bibitem[Auri\`ere(1982)]{aur82} Auri\`ere, M.  1982, \aap,
%    109, 301
%\end{thebibliography}

\clearpage

\end{document}